# Geometric Hybrid Poincaré Sphere with Variable Poles

CHIHIRO TAGO[1], TAKASHI KAKUE[1], AND KEN MORITA[1,*]

[1]*Graduate School of Engineering, Chiba University, Chiba 263-8522, Japan*

[*]*morita@chiba-u.jp*

**Abstract**

We propose a geometric hybrid Poincaré sphere (GHPS) as a unified geometrical framework for describing structured photon states with independently controllable spin angular momentum (SAM) and orbital angular momentum (OAM). Unlike the conventional higher-order Poincaré sphere, in which the SAM and OAM are intrinsically coupled through fixed basis states, the GHPS is constructed by defining its poles as direct products of arbitrary orthogonal bases on the Poincaré sphere (PS) and orbital Poincaré sphere (OPS) and by superposing these pole states. Using numerical simulations, we analyze representative GHPS states and show that the GHPS spherical coordinates govern the amplitude ratio and relative phase between the pole bases. This framework enables spatially inhomogeneous polarization distributions and intensity patterns, including nonseparable structures in which polarization and intensity are intrinsically intertwined, and provides a systematic state-space description for the coherent geometrical control of advanced structured light fields.

## 1. Introduction

The polarization of light represents one of the most fundamental and universal degrees of freedom, corresponding to the spin angular momentum (SAM) of photons. Classically, polarization is described by the amplitude and phase difference between two orthogonal electric-field components transverse to the propagation direction. In quantum mechanics, polarization serves as an intrinsic two-level system of photons, forming the basis of quantum information technologies [1] through superposition and entanglement [2,3]. In contrast, the orbital angular momentum of light is associated with optical vortex beams characterized by a helical phase factor exp ($il\phi$) and a donut-shaped intensity distribution [4–6]. Thus, light possesses not only spin angular momentum (SAM) but also orbital angular momentum (OAM), which is encoded in the spatial structure of the wavefront. Although the SAM represents an *internal* (point-like) degree of freedom, the OAM corresponds to an *external* spatial degree of freedom that can, in principle, take infinitely many values. Such OAM states have attracted increasing interest for large-capacity optical communications [7,8] and high-dimensional quantum information processing [9–11].

To intuitively understand and visualize these SAM and OAM states and their superpositions, scholars have widely used geometrical representations such as the Poincaré sphere (PS) and orbital Poincaré sphere (OPS) [12,13]. For the conventional PS, any polarization state can be expressed as a superposition of two orthogonal circular-polarization bases, whereas the OPS represents the superposition of a pair of OAM modes with opposite topological charges ($\pm l$). Although the high dimensionality of the OAM cannot be fully described by the OPS, this two-dimensional geometric representation effectively captures the relative amplitude and phase relationships of the constituent modes.

A representative extension of this concept is the higher-order Poincaré sphere (HOPS) [14], which integrates SAM and OAM degrees of freedom within a single spherical representation. The HOPS enables the visualization of complex vector beams such as radial and azimuthal polarizations [15], in which polarization and spatial structure are inseparably coupled. Further generalizations, including the hybrid-order Poincaré sphere (HyOPS) [16] and generalized Poincaré sphere (GPS) [17], have been proposed to represent hybrid SAM–OAM coupling states [18,19]. In particular, while arbitrary superpositions of SAM have been considered and geometrically represented in previous study [20], the OAM degree of freedom has almost always been treated as a fixed pair of opposite topological charges. To the best of our knowledge, a fully hybrid framework that allows arbitrary superpositions in both SAM and OAM, on an equal footing, has not yet been established. This suggests that the geometric nature of light has not yet been fully explored, and entirely new classes of structured light may still await discovery.

In this paper, we propose the geometric hybrid Poincaré sphere (GHPS), a unified geometrical framework that can represent arbitrary combinations of SAM and OAM states. The GHPS defines orthogonal basis pairs corresponding to arbitrary antipodal points on both the polarization PS and the OPS, and constructs generalized hybrid states through their tensor product. This representation encompasses the previously reported HOPS [14] and HyOPS [16] as special cases while extending the parameter space to six independent variables: two each from the PS, OPS, and GHPS itself. Thus, the GHPS provides a unified geometric representation for arbitrary spin–orbit coupled states of light and offers a new design principle for structured photons. Furthermore, the GHPS provides a geometric design principle that naturally connects photonic spin–orbit structures with the spin textures available in semiconductor materials. This unified viewpoint encompasses SAM transfer [21,22], helicity-pattern imprinting [23], and the coherent transfer of higher-order polarization states [20], and it guides the development of spin- and mode-selective single-photon–matter interfaces in semiconductor quantum disks [24].

## 2. Theory

The degrees of freedom of structured light are characterized by two physical quantities: the SAM and OAM [4]. In this section, we first outline the concepts of the PS for the SAM and the OPS for the OAM [10], in which each degree of freedom is projected onto a two-dimensional complex vector space and represented geometrically. We then relax the restrictions of the conventional unifying method, the

HOPS [14], and introduce the GHPS, explicitly formulating its construction and advantages, which provide a broader representation space.

### *2.1 Poincaré sphere (PS) and orbital Poincaré sphere (OPS)*

The degrees of freedom associated with a photon's angular momentum, namely the SAM and OAM, can be visually and uniformly represented on geometric spherical spaces called the PS and OPS, respectively (Fig. 1). An arbitrary photon state $|\psi_\xi\rangle$ on the PS or OPS can be generally expressed as a superposition of two orthogonal basis states using the spherical coordinates, latitude $\theta_\xi$ and longitude $\varphi_\xi$, as

$$|\psi_\xi\rangle = \cos\frac{\theta_\xi}{2}|N_\xi\rangle + e^{i\varphi_\xi}\sin\frac{\theta_\xi}{2}|S_\xi\rangle. \qquad (1)$$

Here, $\xi = s$ denotes the SAM state on the PS, and $\xi = o$ denotes the OAM state on the OPS. On the PS, the north and south pole states $|N_s\rangle$ and $|S_s\rangle$ are placed at the right- and left-circular polarization bases $|R\rangle$ and $|L\rangle$, respectively. The PS is described using a set of Stokes parameters $(S_1^{(s)}, S_2^{(s)}, S_3^{(s)})$ defined for the two-dimensional SAM space, as shown in Fig. 1(a). The latitude $\theta_s$ on the PS determines the relative weights of the circular polarization components $|R\rangle$ and $|L\rangle$, thereby defining the ellipticity of the polarization state, while the longitude $\varphi_s$ specifies the relative phase between them and determines the orientation of the polarization state. On the OPS, the north and south pole states $|N_o\rangle$ and $|S_o\rangle$ are placed at two different OAM modes $|l\rangle$ and $|m\rangle$ ($l \neq m$), respectively. The OPS is described using a set of Stokes-like parameters $(S_1^{(o)}, S_2^{(o)}, S_3^{(o)})$ defined for the two-dimensional OAM subspace, as shown in Fig. 1(b). The latitude $\theta_o$ determines the mixing ratio between the OAM components $|l\rangle$ and $|m\rangle$, whereas the longitude $\varphi_o$ provides the relative phase between them, thereby determining the shape of the intensity distribution and the orientation of its spatial structure. The superposition of these two OAM modes produces, in real space, an interference pattern with $|l - m|$ lobes. Consequently, spatial interference structures (intensity patterns) arising from the OAM degree of freedom can be geometrically represented on the OPS.

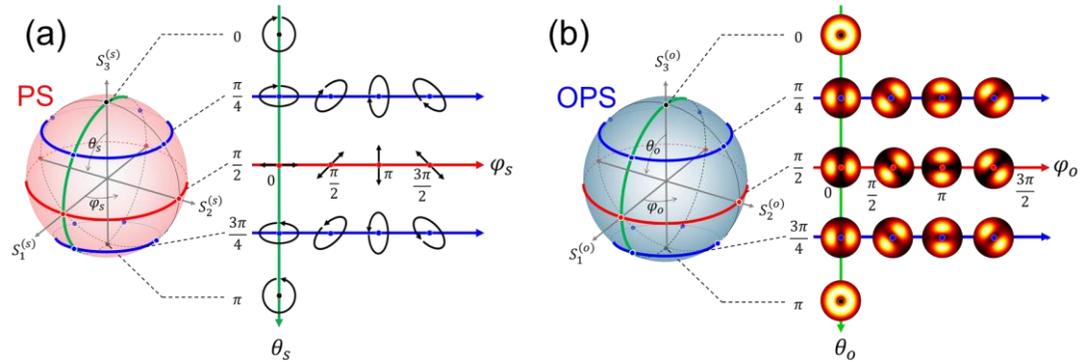

**Fig. 1.** Geometrical representation of photon states on the (a) Poincaré sphere (PS) for spin angular momentum (SAM) and (b) orbital Poincaré sphere (OPS) for orbital angular momentum (OAM).

## 2.2 Higher-order Poincaré sphere (HOPS) and Geometric Hybrid Poincaré Sphere (GHPS)

The HOPS [14] was proposed as the first attempt to integrate SAM and OAM degrees of freedom and represent them on a single sphere [Fig. 2(a)]. This integration is achieved by taking the direct product of the north and south pole states on the PS and OPS and adopting them as the poles of a new sphere. That is, the basis states used in the conventional HOPS are given by

$$|N_H\rangle = |R\rangle \otimes |l\rangle, \qquad |S_H\rangle = |L\rangle \otimes |m\rangle. \qquad (2)$$

An arbitrary photon state $|\psi_H\rangle$ on the HOPS can then be expressed, using the spherical coordinates $\theta_H$ and $\varphi_H$, as a superposition of these pole states:

$$|\psi_H\rangle = \cos\frac{\theta_H}{2}|N_H\rangle + e^{i\varphi_H}\sin\frac{\theta_H}{2}|S_H\rangle. \qquad (3)$$

This representation enables us to geometrically define polarization states that depend on the azimuthal angle, such as vector and radially polarized beams. However, in the HOPS, the SAM and OAM components at the poles $|N_H\rangle$ and $|S_H\rangle$ are fixed to $|R\rangle$, $|L\rangle$ and $|l\rangle$, $|m\rangle$, respectively. Consequently, the two degrees of freedom cannot be controlled independently, and the variety of structured light that can be generated is limited.

To overcome this limitation, we propose the GHPS, which employs arbitrary superposition states on the PS and OPS as hybrid basis states. In the GHPS framework, let $|\psi_\xi^+\rangle$ and $|\psi_\xi^-\rangle$ ($\xi = s, o$) denote arbitrary orthogonal states on the PS and OPS. These states can be expressed in a unified form using the corresponding spherical coordinates $(\theta_\xi, \varphi_\xi)$ on the PS and OPS as

$$|\psi_\xi^+\rangle = \cos\frac{\theta_\xi}{2}|N_\xi\rangle + e^{i\varphi_\xi}\sin\frac{\theta_\xi}{2}|S_\xi\rangle, \qquad (4)$$

$$|\psi_\xi^-\rangle = \sin\frac{\theta_\xi}{2}|N_\xi\rangle - e^{i\varphi_\xi}\cos\frac{\theta_\xi}{2}|S_\xi\rangle. \qquad (5)$$

Subsequently, the "north-pole basis" $|N_G\rangle$ and "south-pole basis" $|S_G\rangle$ of the GHPS are then defined as the direct products of these orthogonal states:

$$|N_G\rangle = |\psi_s^+\rangle \otimes |\psi_o^+\rangle, \qquad |S_G\rangle = |\psi_s^-\rangle \otimes |\psi_o^-\rangle. \qquad (6)$$

Finally, an arbitrary photon state $|\psi_G\rangle$ on the GHPS can be expressed as a superposition of these pole states using the GHPS spherical coordinates $(\theta_G, \varphi_G)$:

$$|\psi_G\rangle = \cos\frac{\theta_G}{2}|N_G\rangle + e^{i\varphi_G}\sin\frac{\theta_G}{2}|S_G\rangle. \qquad (7)$$

Here, $(\theta_G, \varphi_G)$ denote the latitude and longitude on the GHPS, which control the amplitude ratio and relative phase between the $|N_G\rangle$ and $|S_G\rangle$. Unlike the conventional HOPS, the GHPS does not admit a globally fixed Stokes frame. Instead, the Stokes axes $(S_1^{(G)}, S_2^{(G)}, S_3^{(G)})$ are not defined a priori but are determined by the choice of the pole states, thereby forming a moving frame that varies with the GHPS construction. With this formulation, the SAM and OAM degrees of freedom can be treated independently and flexibly. The state is then characterized by six free parameters in total: the PS parameters $(\theta_s, \varphi_s)$, the OPS parameters $(\theta_o, \varphi_o)$, and the GHPS parameters $(\theta_G, \varphi_G)$. This framework enables a comprehensive description of complex structured-light states that simultaneously exhibit spatially inhomogeneous polarization distributions and intricate intensity patterns beyond the scope of conventional HOPS.

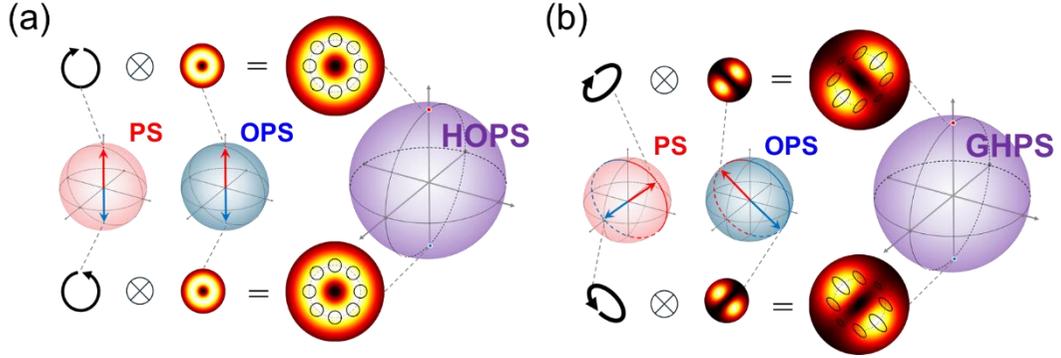

**Fig. 2.** (a) Construction of the hybrid orbital Poincaré sphere (HOPS), obtained from the product of fixed SAM and OAM basis states. The HOPS incorporates only the azimuthal OAM phase degree of freedom, resulting in a fixed donut-shaped radial intensity profile and allowing only rotational variations of the polarization and spatial patterns. (b) Construction of the geometric hybrid Poincaré sphere (GHPS), where both SAM and OAM basis states are coherently mixed with independently tunable parameters $(\theta_s, \varphi_s)$ and $(\theta_o, \varphi_o)$. With its own tunable spherical parameters $(\theta_G, \varphi_G)$, the GHPS enables structural modulation beyond the HOPS, producing rich spatial–polarization textures that result from the full geometrical coupling between the PS and OPS.

## 3. Simulation

In this section, we numerically analyze the beam cross-sectional structures of photon states represented on the GHPS. Our aim is to clarify how the beam structures are determined by the full set of geometric parameters defining the GHPS. As a first step, before discussing general superposition states, we focus on the GHPS pole states and investigate how their beam profiles depend on the latitude parameters of the PS and OPS.

### *3.1 Geometric Hybrid Poincaré Sphere (GHPS) pole states*

The GHPS pole states $|N_G\rangle$ and $|S_G\rangle$ are defined by four independent geometric parameters on the PS and OPS, namely $(\theta_s, \varphi_s, \theta_o, \varphi_o)$. As stated above, the latitude parameters $(\theta_s, \theta_o)$ determine the structural variations of SAM (polarization state) and OAM (intensity distribution), respectively, whereas the longitude parameters $(\varphi_s, \varphi_o)$ determine the principal-axis directions of the polarization and spatial modes, respectively. In particular, the latitude parameters $(\theta_s, \theta_o)$ are the most important, because they do not simply rotate the principal axes but alter the spatial structures themselves, thereby determining the final beam configuration. Therefore, before discussing the GHPS superposition states, we first analyze the two pole states $|N_G\rangle$ and $|S_G\rangle$ and their dependence on the latitude parameters $(\theta_s, \theta_o)$.

Figure 3 illustrates the continuous change in the GHPS pole states as the PS latitude $\theta_s$ (vertical axis) and OPS latitude $\theta_o$ (horizontal axis) are varied. Here, for simplicity, the basis modes are chosen as $l = +1$ and $m = -1$.

1. **Variation in the PS basis (vertical, $\theta_s$)**: When moving along the vertical direction in Fig. 3, corresponding to variations in the PS latitude $\theta_s$, only the polarization state changes continuously while the OAM intensity distribution remains unchanged. This vertical variation directly reflects the state evolution on the PS shown in Fig. 1(a), demonstrating that the GHPS basis faithfully inherits the characteristics of the PS.

2. **Variation in the OPS basis (horizontal, $\theta_o$)**: In contrast, when moving along the horizontal direction in Fig. 3, corresponding to variations in the OPS latitude $\theta_o$, the lobe structure and intensity distribution of the beam change continuously while the SAM polarization state is preserved. The superposition of $l = \pm 1$ shown in Fig. 3 forms a two-lobe structure, since $|l - m| = 2$, as discussed in the previous section. This horizontal variation corresponds to the state evolution on the OPS shown in Fig. 1(b), indicating that the GHPS basis also faithfully inherits the characteristics of the OPS.
3. **Relation between the north- and south-pole bases**: In Fig. 3, the north-pole basis $|N_G\rangle$ and south-pole basis $|S_G\rangle$ exhibit similar features with respect to variations in $\theta_s$ (vertical) and $\theta_o$ (horizontal), but their polarization states are mutually inverted (e.g., right-circular ↔ left-circular), and the phase of the OAM lobe structure is also inverted. This symmetry reflects the orthogonality of the tensor-product bases of the GHPS, $\langle N_G | S_G \rangle = 0$ and demonstrates that the geometric structure of the GHPS is physically consistent.

These results arise from the fact that the GHPS pole bases are rigorously constructed as the tensor product of the PS and the OPS. Owing to this construction, the latitude parameters $(\theta_s, \theta_o)$ function as mutually independent control variables, each governing the structure of the SAM associated with polarization and the intensity-distribution structure of the OAM associated with spatial modes, respectively. Consequently, the GHPS enables a unified geometric description of SAM and OAM while preserving their full independent controllability. Such a property is fundamentally unattainable within the conventional HOPS, where these degrees of freedom are intrinsically coupled, and it therefore constitutes a decisive advantage of the GHPS for structured-light design.

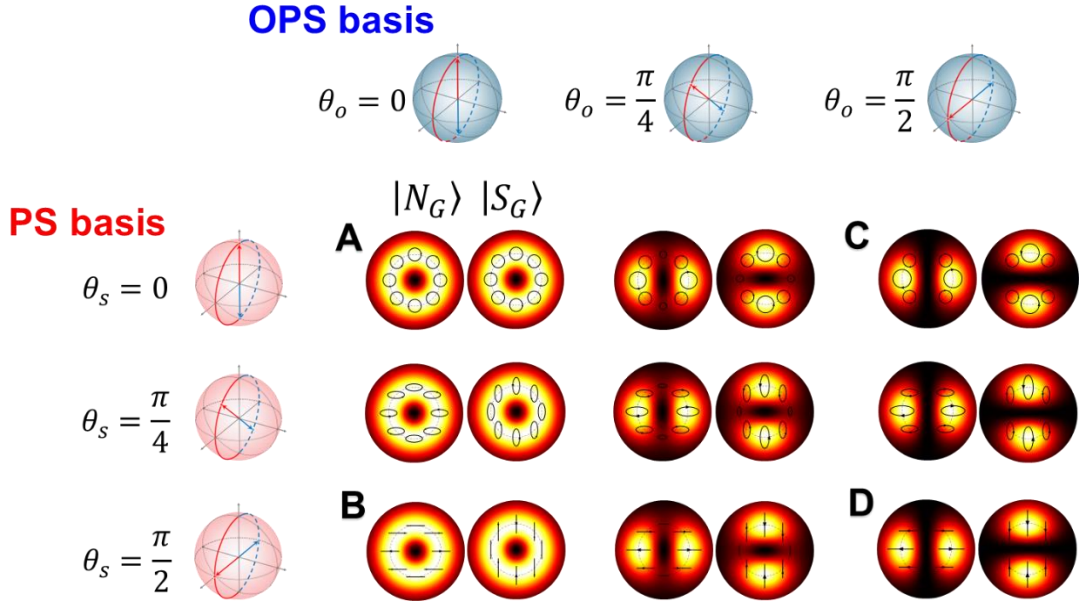

**Fig. 3.** Spatial intensity and local polarization distributions of the GHPS basis states $|N_{so}\rangle$ and $|S_{so}\rangle$ as functions of the PS latitude $\theta_s$ (rows) and OPS latitude $\theta_o$ (columns). The PS latitude $\theta_s$ determines the ellipticity of the polarization, whereas the OPS latitude $\theta_o$ controls the spatial-mode composition of ±1 OAM states. The resulting beam patterns form four representative structural families (A–D), which serve as the basis states for constructing general GHPS superposition states shown in Fig. 4.

## 3.2 Analysis of states on the GHPS sphere

Using the pole states $|N_G\rangle$ and $|S_G\rangle$, which independently control the SAM and OAM degrees of freedom as established in the previous section, we analyze arbitrary photon states on the GHPS by varying the spherical coordinates $(\theta_G, \varphi_G)$. The parameters $\theta_G$ and $\varphi_G$ specify the amplitude ratio and relative phase between the two pole bases, respectively, thereby uniquely determining a superposition state on the GHPS.

Figure 4 shows representative examples of diverse photon states achieved by the six degrees of freedom $(\theta_s, \varphi_s, \theta_o, \varphi_o, \theta_G, \varphi_G)$ on the GHPS. States A–D are classified by specific combinations of the bases (Fig. 3), and their bases are determined by four parameters $(\theta_s, \varphi_s, \theta_o, \varphi_o)$. The common physical origin of the states A–D can be understood directly from the structure of the GHPS pole bases. As defined in Eq. (6), the north- and south-pole bases are constructed as tensor products of the polarization and orbital components, $|N_G\rangle = |\psi_s^+\rangle \otimes |\psi_o^+\rangle$ and $|S_G\rangle = |\psi_s^-\rangle \otimes |\psi_o^-\rangle$. Because the polarization components $|\psi_s^+\rangle$ and $|\psi_s^-\rangle$ are mutually orthogonal by construction, interference between the two pole bases does not appear in intensity measurements. As a result, the intensity distribution is given by the weighted sum of the individual OPS contributions,

$$I(\mathbf{r}) \propto \cos^2\frac{\theta_G}{2}|\psi_o^+(\mathbf{r})|^2 + \sin^2\frac{\theta_G}{2}|\psi_o^-(\mathbf{r})|^2, \quad (8)$$

where the relative weights are controlled by the superposition parameter $\theta_G$.

In contrast to the intensity distribution, the polarization distribution originates from the coherent superposition of the two orthogonal polarization components $|\psi_s^+\rangle$ and $|\psi_s^-\rangle$. At a given azimuthal angle $\phi$, the local polarization state can be expressed as

$$|\psi_s(\phi)\rangle \propto |\psi_s^+\rangle + \rho(\phi)e^{i\delta(\phi)}|\psi_s^-\rangle, \quad (9)$$

where the relative phase $\delta(\phi) = \varphi_G + \arg[\psi_o^+(\phi)] - \arg[\psi_o^-(\phi)]$ depends on the azimuthal angle through the orbital component as well as on the GHPS longitude parameter $\varphi_G$, while $\rho(\phi) = \tan(\theta_G/2)|\psi_o^+(\phi)|/|\psi_o^-(\phi)|$ represents the local amplitude ratio determined by both the superposition parameter $\theta_G$ and the OPS intensity profiles. Here, we assume that the OPS basis states share the same radial envelope, so that the spatial dependence is captured primarily by the azimuthal angle $\phi$. As $\phi$ varies, the phase $\delta(\phi)$ changes continuously, leading to an azimuthally varying local polarization state. This behavior can be intuitively understood as a rotation of the polarization state on the PS, as illustrated in Fig. 1(a).

1. **State A.** The pole bases are constructed from the direct product of the PS and OPS poles, $(e^{i\phi}|R\rangle, e^{-i\phi}|L\rangle)$, which corresponds to the conventional HOPS states. The overall intensity distribution $I(\mathbf{r})$ given by Eq. (8) remains azimuthally uniform, because the contributing OAM modes ($e^{\pm i\phi}$) share the same ring-shaped intensity profile. As a result, the donut-shaped intensity distribution is preserved. The polarization behavior follows directly from Eq. (9) and is governed by a simple azimuthal phase dependence $\delta(\phi) = 2\phi$. As the latitude parameter $\theta_G$ varies, the polarization ellipticity changes uniformly. On the equator, the equal superposition of opposite circular polarizations yields linear polarization at all azimuthal angles $\phi$. Well-known vector beams, such as radial and azimuthal polarizations, are included as special cases. The longitude parameter $\varphi_G$ tilts the polarization direction at each azimuthal angle, in direct analogy with rotations in the $S_1^{(s)}$–$S_2^{(s)}$ plane of the PS [Fig. 1(a)].
2. **State B.** The pole bases are constructed from the direct product of linear-polarization-based PS bases and OPS poles, $(e^{i\phi}|H\rangle, e^{-i\phi}|V\rangle)$. The overall intensity distribution $I(\mathbf{r})$ given by Eq. (8) remains azimuthally uniform, for the same reason as in state A. As $\theta_G$ varies, regions with azimuthally varying polarization ellipticity coexist with regions where the ellipticity remains unchanged. On the equator, the state becomes an equal superposition, and the local polarization $|\psi_s(\phi)\rangle$ is determined by the azimuthal phase difference $\delta(\phi) = 2\phi$, resulting in alternating linear, elliptical, and circular polarizations. This behavior corresponds to a rotation of the local polarization state in the $S_2^{(s)}$–$S_3^{(s)}$ plane of the PS [Fig. 1(a)].

3. **State C.** The pole bases are constructed from the direct product of the PS poles and the OPS equatorial bases, $((e^{i\phi} + e^{-i\phi})|R\rangle, (e^{i\phi} - e^{-i\phi})|L\rangle)$. Since these OPS bases lie on the equator of the OPS in Fig. 1(b), the intensity distribution given by Eq. (8) is obtained as an incoherent superposition of two orthogonal two-lobe intensity patterns corresponding to the OPS basis states. The superposition parameter $\theta_G$ therefore controls the relative weights of the two OPS contributions and hence the contrast of the intensity lobes. At mid-latitudes, the intensity is dominated by the contribution of the nearer pole, whereas on the equator the equal superposition yields a uniform intensity distribution. The polarization behavior follows from Eq. (9) with a more complex azimuthal phase dependence than in state A. Owing to the phase factors ($e^{i\phi} \pm e^{-i\phi}$), the relative phase $\delta(\phi)$ leads to linear polarization at diagonal azimuthal angles $\phi = \pi/4, 3\pi/4, 5\pi/4, 7\pi/4$, while circular polarization appears at horizontal and vertical directions $\phi = 0, \pi/2, \pi, 3\pi/2, 2\pi$. Consequently, linear and circular polarizations appear alternately along the azimuthal direction, similarly to the equatorial state of state B. At mid-latitudes, the unequal weighting of the two pole bases prevents the formation of purely linear polarization, resulting instead in a mixture of circular and elliptical polarization components accompanied by azimuthally varying intensity.

4. **State D.** The pole bases are constructed from the direct product of the PS and OPS equatorial bases $((e^{i\phi} + e^{-i\phi})|H\rangle, (e^{i\phi} - e^{-i\phi})|V\rangle)$. This most extended combination of pole bases incorporates the characteristics of both state B, which employs extended PS bases, and state C, which employs extended OPS bases. The resulting intensity distribution is identical to that of state C. The polarization behavior follows from Eq. (9) with a phase difference $\delta(\phi)$. Owing to the characteristic intensity extrema of the OPS contributions, the polarization states at specific azimuthal angles ($\phi = 0, \pi/4, \pi/2, 3\pi/4$) are fixed to linear polarization in the superposed state. At the remaining azimuthal positions on the equator, the local polarization state varies according to the phase difference $\delta(\phi)$, resulting in alternating linear, elliptical, and circular polarizations. This enables rich polarization combinations based on linear-polarization components within the GHPS framework.

A more quantitative characterization can be obtained by explicitly substituting the PS and OPS basis functions into Eqs. (8) and (9). However, the qualitative analysis presented here already captures the essential physical mechanisms governing the GHPS states. These results clearly demonstrate that the polarization and intensity distributions form a nonseparable, intertwined structure, and that the six degrees of freedom of the GHPS enable an exceptionally high level of spatial structuring.

Although an experimental demonstration of the GHPS states is beyond the scope of this paper, we briefly comment on their experimental feasibility. GHPS states represent spatially structured light fields in which phase and intensity are jointly controlled. Such control can be efficiently implemented on a spatial light modulator using phase-only modulation based on the dual-phase representation of complex fields [25,26] and realized in practice via double-phase encoding techniques [27]. This approach enables independent control of the orthogonal OPS basis states and the realization of the proposed GHPS framework without requiring amplitude-modulating devices.

Looking ahead, the GHPS offers a unified geometric platform for designing structured photons with coupled spin and orbital degrees of freedom. Recent progress in high-dimensional quantum-structured light [28] and real-space topological polarization textures, such as optical skyrmions [29], highlights future directions in which systematic state-space descriptions may become increasingly important. In this sense, the GHPS may provide a useful bridge between abstract geometric representations and experimentally realizable structured light fields, opening new avenues for photonic state engineering and light–matter interfaces.

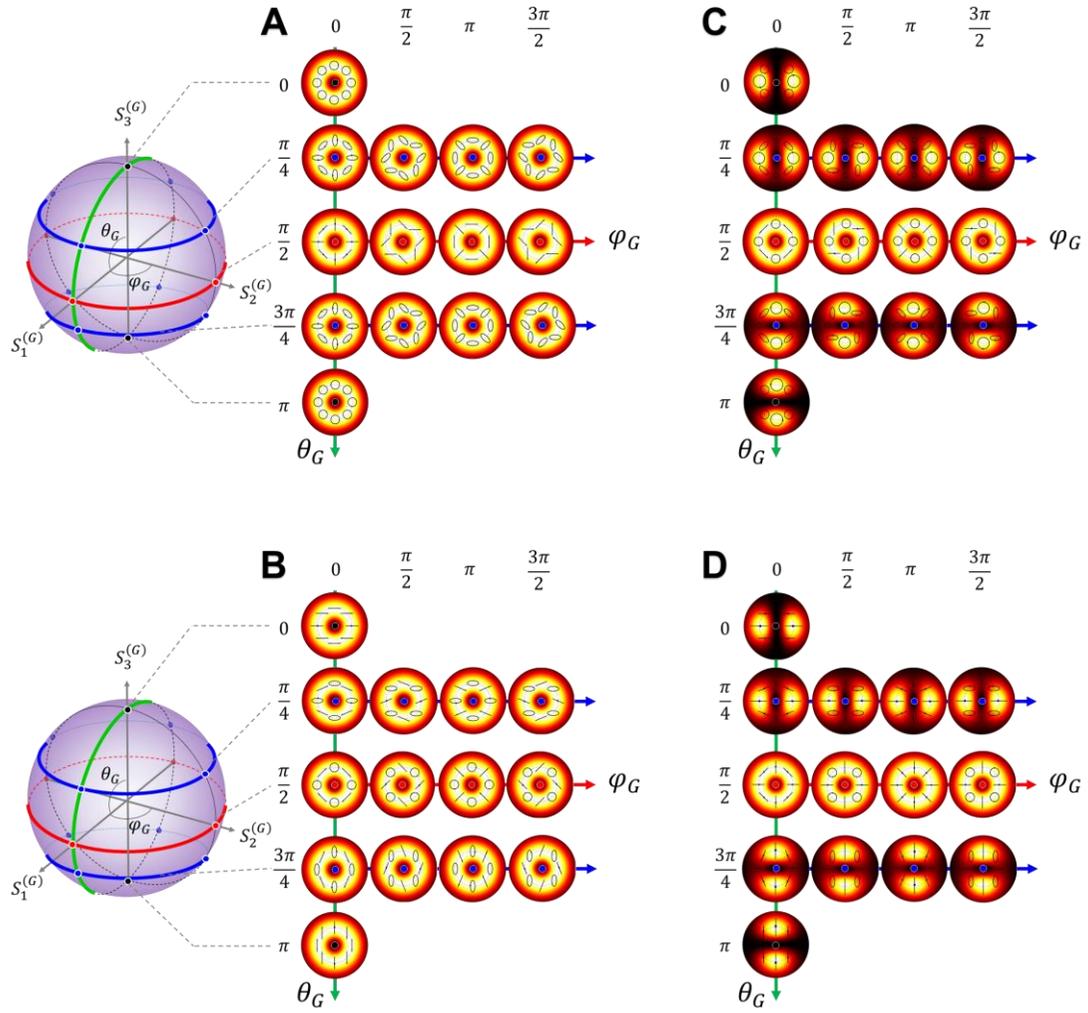

**Fig. 4.** Geometrical mapping of general GHPS states for the four structural families (A–D). Each panel shows the spatial intensity and polarization textures of hybrid states parameterized by the GHPS spherical coordinates $(\theta_G, \varphi_G)$. The polar angle $\theta_G$ modulates the intrinsic spatial structure (nodal arrangement, ring formation, or central-mode features), whereas the azimuthal angle $\varphi_G$ induces rotational or translational modulation of the pattern. Together, these maps demonstrate that GHPS defines a full two-dimensional manifold of spin–orbit structured light, extending beyond the descriptive capability of the conventional HOPS.

## 4. Conclusion

We have proposed a generalized hybrid Poincaré sphere (GHPS) in which the north and south poles are defined by the direct products of two mutually orthogonal superposition states of spin angular momentum (SAM) and orbital angular momentum (OAM). The GHPS provides a unified and higher-order state-space framework that encompasses the conventional Poincaré sphere (PS), the orbital Poincaré sphere (OPS), and their hybrid extensions, such as the hybrid Poincaré sphere (HOPS), while enabling a more general description of intrinsically nonseparable SAM–OAM states. Within this framework, we have identified previously unexplored optical modes exhibiting simultaneous spatial modulation of polarization and intensity, arising from the structured interplay between SAM and OAM. Thus, the GHPS extends beyond a descriptive representation of structured light fields, providing a unified theoretical platform for designing advanced spin–orbit structured light and for bridging photonic state geometry with spin–orbit textures in solid-state systems.

**Funding.** Japan Society for the Promotion of Science (22H01981, 25H01608, 25K01687); Murata Science Foundation (AN21259); JKA Foundation (2024M-374); Toshiba Devices & Storage Academic Incentive System 2024, 2025.

**Acknowledgments.** We would like to thank Editage (www.editage.jp) for their English language editing.

**Disclosures.** The authors declare no conflicts of interest.

**Data availability.** The data underlying the results presented in this paper are not publicly available at this time but may be obtained from the authors upon reasonable request.